# Finding Quasars behind the Galactic Plane. II. Spectroscopic Identifications of 204 Quasars at $|b| < 20°$

Yuming Fu[1,2], Xue-Bing Wu[1,2], Linhua Jiang[1,2], Yanxia Zhang[3], Zhi-Ying Huo[4], Y. L. Ai[5,6], Qian Yang[7], Qinchun Ma[1,2], Xiaotong Feng[1,2], Ravi Joshi[2], Wei Jeat Hon[8], Christian Wolf[9,10], Jiang-Tao Li[11,12], Jun-Jie Jin[1,2], Su Yao[13], Yuxuan Pang[1,2], Jian-Guo Wang[14,15], Kai-Xing Lu[14,15], Chuan-Jun Wang[14,15], Jie Zheng[3], Liang Xu[14,15], Xiao-Guang Yu[14,15,16], Bao-Li Lun[14,15], and Pei Zuo[2,17]

[1] Department of Astronomy, School of Physics, Peking University, Beijing 100871, People's Republic of China; wuxb@pku.edu.cn, fuym@pku.edu.cn
[2] Kavli Institute for Astronomy and Astrophysics, Peking University, Beijing 100871, People's Republic of China
[3] CAS Key Laboratory of Optical Astronomy, National Astronomical Observatories, Beijing 100101, People's Republic of China
[4] National Astronomical Observatories, Chinese Academy of Sciences, Beijing 100101, People's Republic of China
[5] College of Engineering Physics, Shenzhen Technology University, Shenzhen 518118, People's Republic of China
[6] Shenzhen Key Laboratory of Ultraintense Laser and Advanced Material Technology, Shenzhen 518118, People's Republic of China
[7] Department of Astronomy, University of Illinois at Urbana-Champaign, Urbana, IL 61801, USA
[8] School of Physics, University of Melbourne, Parkville, Victoria 3010, Australia
[9] Research School of Astronomy and Astrophysics, Australian National University, Canberra, ACT 2611, Australia
[10] Centre for Gravitational Astrophysics, Australian National University, Canberra, ACT 2600, Australia
[11] Department of Astronomy, University of Michigan, 311 West Hall, 1085 S. University Avenue, Ann Arbor, MI, 48109-1107, USA
[12] Purple Mountain Observatory, Chinese Academy of Sciences, 10 Yuanhua Road, Nanjing 210023, People's Republic of China
[13] Max-Planck-Institut für Radioastronomie, Auf dem Hügel 69, D-53121 Bonn, Germany
[14] Yunnan Observatories, Chinese Academy of Sciences, Kunming 650216, People's Republic of China
[15] Key Laboratory for the Structure and Evolution of Celestial Objects, Chinese Academy of Sciences, Kunming 650216, People's Republic of China
[16] University of Chinese Academy of Sciences, Beijing 100049, People's Republic of China
[17] International Centre for Radio Astronomy Research (ICRAR), University of Western Australia, 35 Stirling Highway, Crawley, WA 6009, Australia
Received 2022 March 14; revised 2022 May 24; accepted 2022 June 10; published 2022 July 29

## Abstract

Quasars behind the Galactic plane (GPQs) are important astrometric references and valuable probes of Galactic gas, yet the search for GPQs is difficult due to severe extinction and source crowding in the Galactic plane. In this paper, we present a sample of 204 spectroscopically confirmed GPQs at $|b| < 20°$, 191 of which are new discoveries. This GPQ sample covers a wide redshift range from 0.069 to 4.487. For the subset of 230 observed GPQ candidates, the lower limit of the purity of quasars is 85.2%, and the lower limit of the fraction of stellar contaminants is 6.1%. Using a multicomponent spectral fitting, we measure the emission line and continuum flux of the GPQs, and estimate their single-epoch virial black hole masses. Due to selection effects raised from Galactic extinction and target magnitude, these GPQs have higher black hole masses and continuum luminosities in comparison to the SDSS DR7 quasar sample. The spectral-fitting results and black hole mass estimates are compiled into a main spectral catalog, and an extended spectral catalog of GPQs. The successful identifications prove the reliability of both our GPQ selection methods and the GPQ candidate catalog, shedding light on the astrometric and astrophysical programs that make use of a large sample of GPQs in the future.

*Unified Astronomy Thesaurus concepts:* Active galactic nuclei (16); Catalogs (205); Galactic and extragalactic astronomy (563); Quasars (1319); Spectroscopy (1558); Supermassive black holes (1663)

*Supporting material:* figure set, machine-readable tables

## 1. Introduction

Quasars are active galactic nuclei (AGNs) of very high luminosity, found over a wide range of redshifts. So far, the most distant known quasar is J0313-1806 at redshift $z = 7.642$ (Wang et al. 2021). Quasars are key to understanding the formation and evolution of supermassive black holes and their host galaxies (e.g., Di Matteo et al. 2005; Kormendy & Ho 2013), probes of the interstellar and intergalactic mediums at different redshifts (e.g., Weymann et al. 1981; Rees 1986; Trump et al. 2006), and building blocks of large-scale structures of the universe (e.g., Eisenstein et al. 2011; Dawson et al. 2013; Blanton et al. 2017). Located at cosmic distances and being point sources with small parallaxes and proper motions, quasars are also ideal references for astrometry, with which celestial reference frames can be defined (e.g., Ma et al. 2009; Mignard et al. 2016; Gaia Collaboration et al. 2018, 2022).

Although the population of known quasars has grown significantly with the help of surveys in various bands over the past few decades, the number of quasars behind the Galactic plane (GPQs) remains limited. For example, the 16th data release of the Sloan Digital Sky Survey Quasar Catalog (SDSS DR16Q; Lyke et al. 2020) contains 750,414 spectroscopically identified quasars, but only 3737 (~0.5%) of them are located at $|b| < 20°$. Also, the candidate quasar samples that are used to define the Gaia celestial reference frames only contain a small fraction of sources in the middle Galactic plane. For example, only 5.8% of the sources are located at $|b| < 15°$ (25.9% of the entire sky) in the `agn_cross_id` table (quasar candidates) of the Gaia Early Data Release 3 (Gaia Collaboration et al. 2021; Lindegren et al. 2021) archive.

**Table 1**
Summary of Observed GPQ Candidates

| Telescope | $N_{candidate}$ | $N_{QSO}$ | $N_{star}$ | $N_{galaxy}$ | $N_{unknown}$ | Min($z$) | Max($z$) | Min($i_{P1}$)[a] | Max($i_{P1}$) |
|---|---|---|---|---|---|---|---|---|---|
| LJT | 83 | 76 | 3 | 0 | 4 | 0.099 | 3.744 | 15.69 | 21.00 |
| XLT | 76 | 61 | 3 | 2 | 10 | 0.071 | 3.541 | 15.51 | 18.68 |
| P200 | 40 | 26 | 8 | 0 | 6 | 0.135 | 4.487 | 17.59 | 19.90 |
| MDM13 | 31 | 28 | 3 | 0 | 0 | 0.069 | 2.337 | 15.80 | 18.15 |
| ANU23 | 13 | 13 | 0 | 0 | 0 | 0.794 | 2.112 | 16.74 | 18.43 |
| Total | 243 | 204 | 17 | 2 | 20 | 0.069 | 4.487 | 15.51 | 21.00 |

**Note.**
[a] The Pan-STARRS1 $i_{P1}$ magnitudes are not extinction corrected.

A large sample of GPQs can help build a better celestial reference frame by directly covering the sky in the Galactic plane, and gain a better knowledge of the systematic astrometry errors of Gaia in the Galactic plane (see, e.g., Arenou et al. 2018; Lindegren et al. 2021; Fabricius et al. 2021; Gaia Collaboration et al. 2022). In addition, absorption lines in the spectra of GPQs can help probe the gas structures of the Milky Way using techniques similar to those of Savage et al. (1993, 2000) and Ben Bekhti et al. (2008, 2012).

A spectral survey for quasars requires careful candidate selections prior to the observations, which have always been difficult in dense stellar fields. A few attempts have explored the feasibility of identifying quasars in such crowded fields. For example, Im et al. (2007) discovered 40 bright quasars at $|b| \leqslant 20°$ out of 601 candidates that are selected with the near-IR color cut of $J - K > 1.4$ in the Two Micron All Sky Survey (Skrutskie et al. 2006) and the detection of a radio counterpart from the NRAO Very Large Array Sky Survey (Condon et al. 1998). Kozłowski & Kochanek (2009) identified 5000 AGNs behind the Magellanic Clouds with the mid-IR color cuts modified from the method of Stern et al. (2005). Huo et al. (2010, 2013, 2015) discovered 1870 new quasars around the Andromeda (M31) and Triangulum (M33) galaxies, with the Large Sky Area Multi-Object Fiber Spectroscopic Telescope (LAMOST, also called the Guoshoujing Telescope; Cui et al. 2012) from 2009 to 2013.

In the first paper of this series (Fu et al. 2021, hereafter Paper I), we reported the selection methods for GPQ candidates based on transfer learning, and a catalog of 160,946 GPQ candidates selected from Pan-STARRS1 (PS1; Chambers et al. 2016) DR1 and the Wide-field Infrared Survey Explorer final catalog release (AllWISE; Wright et al. 2010; Mainzer et al. 2011). As has been evaluated by Paper I, the purity of quasars on the GPQ candidates that are matched to the SIMBAD[18] (Wenger et al. 2000) database is ~90%. Nevertheless, spectroscopic identifications for these GPQ candidates are needed to validate the efficiency of the selection methods, and the precision of the photometric redshifts. In addition, a clean spectroscopic sample of GPQs is required to extend the list of astrometric references in the Galactic observations for our GPQ candidates have been carried out since 2018 with several telescopes. We visually inspect all reduced spectra, and measure the redshifts for identified quasars. In order to quantify the physical properties of GPQs that we discover, we fit their spectra with multicomponent models, and derive quantities including continuum luminosities and black hole masses.

This paper is organized as follows. In Section 2, we introduce the details of the observations for the GPQ candidates. In Section 3, we describe the methods for visual inspections and spectral fitting. In Section 4, we present the results of spectral analysis, and statistical properties of the new GPQ sample as compared to SDSS DR7Q. We summarize the paper in Section 5. Throughout this paper we adopt a flat $\Lambda$CDM cosmology with $\Omega_\Lambda = 0.7$ and $h = 0.7$.

## 2. Observations

Spectroscopic observations for GPQ candidates have been carried out with five telescopes in China, USA, and Australia since 2018 (see Table 1). A total of 243 GPQ candidates have been observed until 2021 May, 230 of which are selected from the GPQ candidate catalog in Paper I. The other 13 targets are selected from the preliminary versions of the GPQ candidate catalog and are not included in the published version (v1.0) in Paper I.

### 2.1. The Xinglong 2.16 m Telescope

The Xinglong 2.16 m Telescope (hereafter XLT; Fan et al. 2016) is an equatorial mount reflector at Xinglong Observatory, National Astronomical Observatories, Chinese Academy of Sciences. Since 2018, we have been using the Beijing Faint Object Spectrograph and Camera on XLT to identify GPQ candidates with low-resolution spectroscopy. The G4 grism (reciprocal linear dispersion 198 Å mm$^{-1}$) and a slit with a width of 2″.3 are used such that the wavelength coverage is 3850–7000 Å, and the spectral resolution ($R = \lambda/\Delta\lambda$) is 265 at 5007 Å (Fan et al. 2016).

### 2.2. The Lijiang 2.4 m Telescope

The Lijiang 2.4 m Telescope (hereafter LJT; Wang et al. 2019) is an altitude-azimuth mount reflector operated by Lijiang Observatory, Yunnan Observatories, Chinese Academy of Sciences. The instrument we use is the Yunnan Faint Object Spectrograph and Camera on LJT. Our typical configuration for observation is the G3 grism and a slit with a width of 2″.5, which can achieve a wavelength coverage of 3400–9100 Å and a spectral resolution of $R \sim 250$ at 6030 Å.

### 2.3. The 200 inch Hale Telescope

The 200 inch (5.3 m) Hale Telescope (P200) is operated by Palomar Observatory. The Double Spectrograph (DBSP; Oke & Gunn 1982) has been used for the observations. DBSP uses a dichroic to split light into red and blue channels (sides), which are observed simultaneously. We use a slit with a width of 1″.5, with the 600 lines mm$^{-1}$ grating for the blue channel (wavelength range 2500–5700 Å), and the 316 lines mm$^{-1}$ grating for the red channel (wavelength range 4800–10700 Å). Such a combination of slit and gratings gives a spectral resolution of $R \sim 900$ at central wavelengths for both channels.

---
[18] http://simbad.cds.unistra.fr/simbad/

### 2.4. The ANU 2.3 m Telescope

The 2.3 m telescope (hereafter ANU23) at Siding Spring Observatory is operated by the Australian National University. For all observations, we use the Wide-Field Spectrograph (Dopita et al. 2007, 2010), which is a double-beam, image-slicing, integral-field spectrograph. The B3000/R3000 grating ($R \sim 3000$) is used for obtaining the blue/red sides of the spectra, with the RT560 beam splitter that splits each spectrum into two parts at 5600 Å. The wavelength range is about 3300–9000 Å.

### 2.5. The McGraw-Hill 1.3 m Telescope

The McGraw-Hill 1.3 m Telescope (hereafter MDM13) is operated by the MDM Observatory, which is located adjacent to the Kitt Peak National Observatory. We use the Boller and Chivens CCD Spectrograph (see Abrahams & Winans 2013) with the combination of the 150 grooves mm$^{-1}$ grating and the 1″ slit, which achieves a resolution of $\sim 4.4$ Å pixel$^{-1}$.

## 3. Spectral Analysis

The spectral data are reduced with the standard IRAF (Tody 1986, 1993) routines. In particular, data from XLT and LJT are reduced with PyFOSC (Fu 2020), a pipeline toolbox based on PyRAF (Science Software Branch at STScI 2012) for the long-slit spectroscopy. All the extracted spectra are visually inspected for source identifications and redshift measurements. For sources that are identified as quasars, we fit their spectra and obtain useful parameters.

### 3.1. Visual Identifications and Redshift Measurements

The visual inspections and redshift measurements are conducted with ASERA (Yuan et al. 2013), a semiautomated graphical user interface toolkit for spectral classification. During the inspection process, both the target spectrum and a template spectrum (of quasars by default) will be displayed for recognition. Template spectra of different types of quasars, galaxies, and stars are embedded in the toolkit for convenience. The user can determine the most likely spectral type for the target, and retrieve the redshift of the target by simply moving the template spectrum along the wavelength axis to match the target spectrum.

From the whole sample of 243 spectra, we identify 204 quasars, 2 galaxies, and 18 stars. The other 19 sources are marked as "unknown" due to poor data quality. Therefore the success rate of the identification is 84%. A summary of different classes, redshift ranges, and magnitude ranges of the observed GPQ candidates is listed in Table 1. As can be seen from Table 1, the efficiencies for GPQ identification of LJT, MDM13, and ANU23 are all above 90%, which are higher than those of XLT (80%) and P200 (65%). Among the five telescopes, MDM13 is mainly used to observe bright targets due to the small aperture, ANU23 is used to identify sources in the southern sky thanks to its location at the Southern Hemisphere, and P200 is mainly used to observe high-$z$ GPQ candidates. The lower efficiency for P200 is mostly due to the higher fraction of faint sources in its target list.

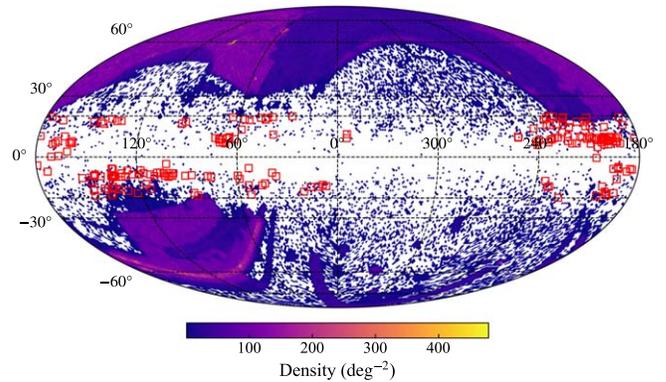

**Figure 1.** The sky distribution of the 204 spectroscopically identified GPQs (red squares) in Galactic coordinates. A HEALPix (Górski et al. 2005) sky density map of known Type 1 QSOs and AGNs from MILLIQUAS v7.4d is shown in the background, with the parameter $N_{\rm side} = 64$ and an area of 0.839 deg$^2$ per pixel.

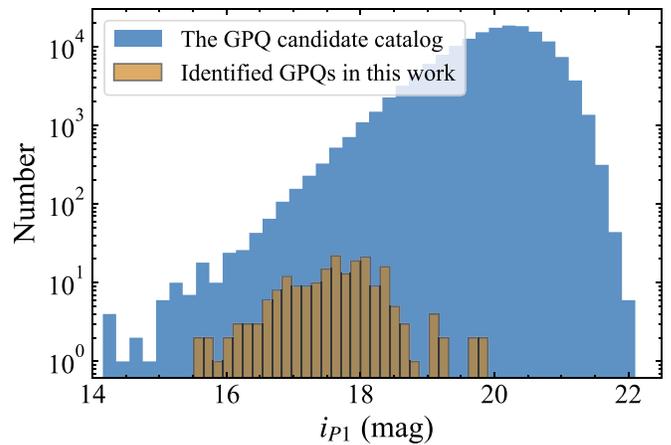

**Figure 2.** Log-scale histograms of the apparent $i_{P1}$ magnitudes of our identified GPQs (orange bars) and the whole GPQ candidate catalog (filled blue step plot).

Crossmatching the 204 GPQs with the million quasar catalog (MILLIQUAS v7.4d; Flesch 2021) results in 13 known quasars/AGNs (labeled with "Q" or "A" in the "Type" column of MILLIQUAS), leaving 191 newly discovered GPQs. The sky distribution of the 204 identified GPQs is shown in Figure 1. The median apparent $i_{P1}$ magnitude of our identified GPQs (17.67) is brighter than that of the GPQ candidate catalog (20.08) by 2.41 mag, which indicates that the identified GPQs are a bright subset of the whole GPQ sample (see also Figure 2). The faintest identified GPQ has $i_{P1}$ of 19.90 mag.

The GPQ candidate catalog shows an $i_{P1}$-band magnitude distribution (Figure 2) that is similar to those of some earlier quasar candidate samples (e.g., Figure 2 of Richards et al. 2009). In the GPQ candidate catalog, only 64 sources are brighter than 16 mag in the $i_{P1}$ band, 18 of which have already been identified as quasars/AGNs. At $i_{P1} < 16$, our observations reveal four new GPQs, as well as four stellar contaminants. Assuming such an efficiency of $\sim$50%, we still expect that the contamination rate from bright stars is lower than 36% in the bright end ($i_{P1} < 16$) for our GPQ candidate sample.

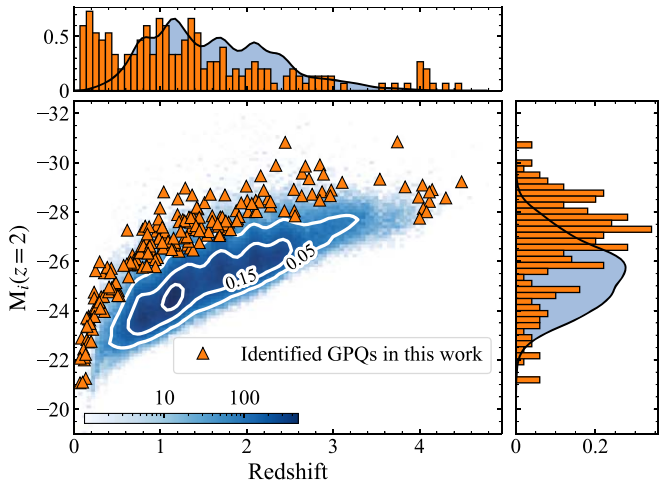

**Figure 3.** The absolute magnitudes $M_i(z=2)$ and redshift distributions of our identified GPQs and the whole GPQ candidate catalog. In the main panel (lower left), our identified GPQs are represented by orange triangles, and the two-dimensional histogram (density plot) of the GPQ candidate catalog is shown in the background. White contour lines based on two-dimensional kernel density estimation (KDE) are displayed on the density plot. In the top and right panels, the blue shaded areas denote the KDE probability density functions of the GPQ candidate catalog, and the orange bars denote the probability densities of our identified GPQs.

To compare the intrinsic brightness of our identified GPQs and the whole GPQ candidate catalog, we calculate the SDSS $i$-band absolute magnitude $M_i$ normalized at $z = 2$ of the two samples. Because SDSS photometry is unavailable for most of our identified GPQs and GPQ candidates, we first convert the $i_{P1}$ magnitude to the $i_{SDSS}$ magnitude with the transformations from Tonry et al. (2012). Then we correct for Galactic extinction for the converted $i_{SDSS}$ with the dust map from Planck Collaboration et al. (2014) (hereafter Planck14) and the extinction law from Wang & Chen (2019). The absolute magnitudes $M_i(z=2)$ are calculated with the K-correction (see e.g., Oke & Sandage 1968; Hogg et al. 2002; Blanton & Roweis 2007) values for the SDSS $i$ band from Richards et al. (2006).

The absolute magnitudes $M_i(z=2)$ and redshift distributions of our identified GPQs and the whole GPQ candidate catalog are shown in Figure 3. In general, the 204 identified GPQs are distributed at the brighter end of the whole GPQ candidate sample, indicating a selection bias toward bright sources of our spectroscopic observations. Such a selection bias is also seen as the high-probability densities of low-redshift ($z \lesssim 0.6$) quasars in the identified sample. The highest redshift of new GPQs that we have reached so far is 4.487. Currently, fewer GPQs are seen at $3 < z < 4$ than at $z > 4$ in our spectral sample, because $z > 4$ targets are preferred in the high-$z$ observations with P200 while bright (mainly $z < 3$) targets are preferred in the observations of other telescopes.

In the above process of calculating absolute magnitudes, the conversions between the PS1 photometric system and SDSS are computed from stellar spectral energy distributions (SEDs), which might be less accurate for quasar SEDs (Tonry et al. 2012). In addition, photometric redshift values are used for the GPQ candidates because the spectral redshifts are unavailable. Such uncertainties have subtle effects on our analysis of the selection biases because the absolute magnitudes of the two samples are calculated in a consistent way.

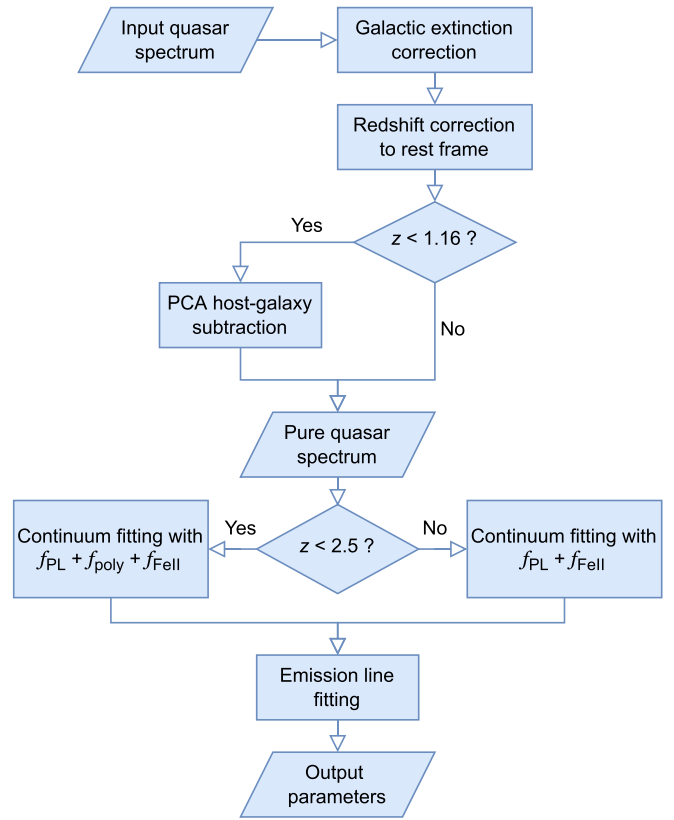

**Figure 4.** Flowchart of the spectral-fitting process with PyQSOFit and QSOFITMORE.

**Table 2**
Line-fitting Parameters

| Line Complex | Fitting Range (Å) | Line | $n_{Gauss}$ |
|---|---|---|---|
| Hα | 6400–6800 | Hα broad | 3 |
| | | Hα narrow | 1 |
| | | [N II]6549 | 1 |
| | | [N II]6585 | 1 |
| | | [S II]6718 | 1 |
| | | [S II]6732 | 1 |
| Hβ | 4640–5100 | Hβ broad | 3 |
| | | Hβ narrow | 1 |
| | | [O III]4959 core | 1 |
| | | [O III]4959 wing | 1 |
| | | [O III]5007 core | 1 |
| | | [O III]5007 wing | 1 |
| Mg II | 2700–2900 | Mg II broad | 2 |
| | | Mg II narrow | 1 |
| C III] | 1850–1970 | C III] | 2 |
| C IV | 1500–1600 | C IV | 3 |

### 3.2. Spectral Fitting with PyQSOFit and QSOFITMORE

To quantify the statistical properties of our newly identified GPQ sample, we develop and utilize QSOFITMORE (version 1.1.0; Fu 2021), a wrapper package based on PyQSOFit (Guo et al. 2018), to fit the 204 spectra of quasars. The original PyQSOFit code is designed for spectra from SDSS instead of those obtained with other facilities. In order to adapt the fitting code to our GPQ spectra, a few new features over PyQSOFit are added to QSOFITMORE, including: (i) input/output supports for spectra in custom (non-SDSS) formats, (ii) an alternative dust map and extinction law for dereddening, and (iii) automatic derivation and output of narrow-line properties (e.g., FWHM, equivalent width, line flux). We refer to Shen et al. (2011, 2019) and Rakshit et al. (2020) for details of the fitting procedure.

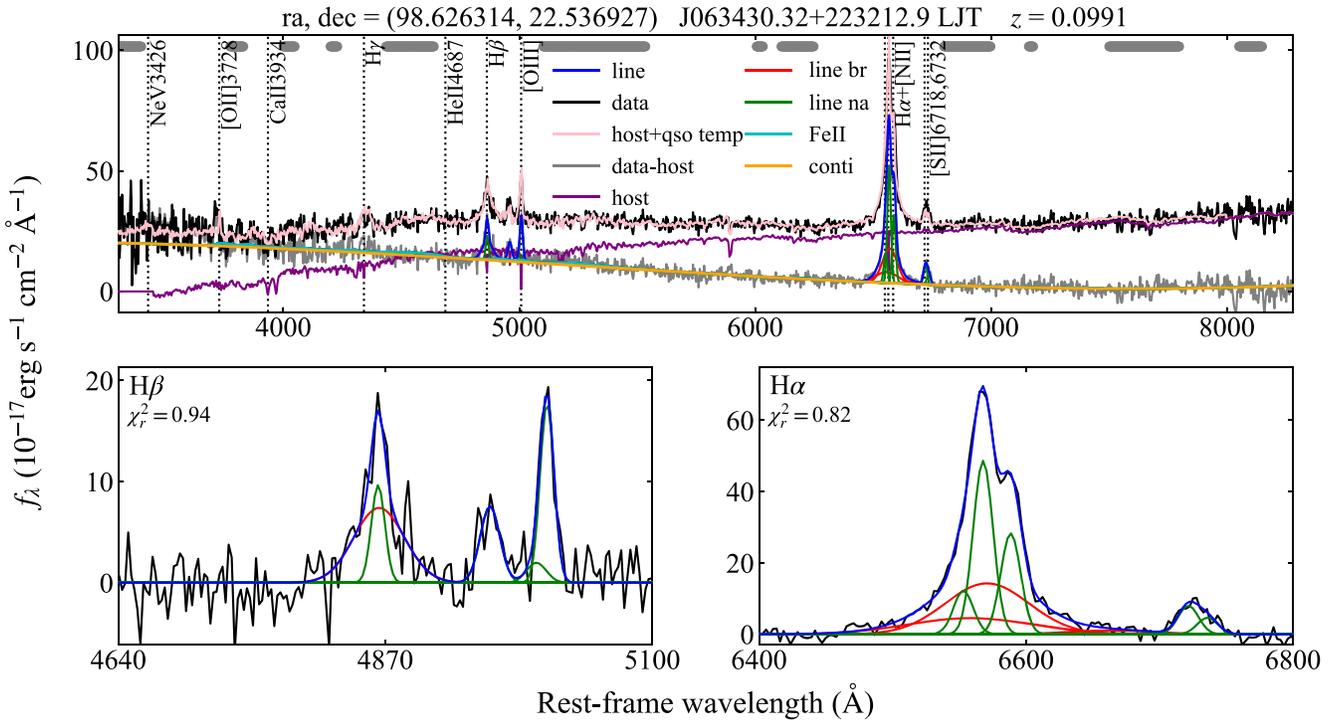

(a)

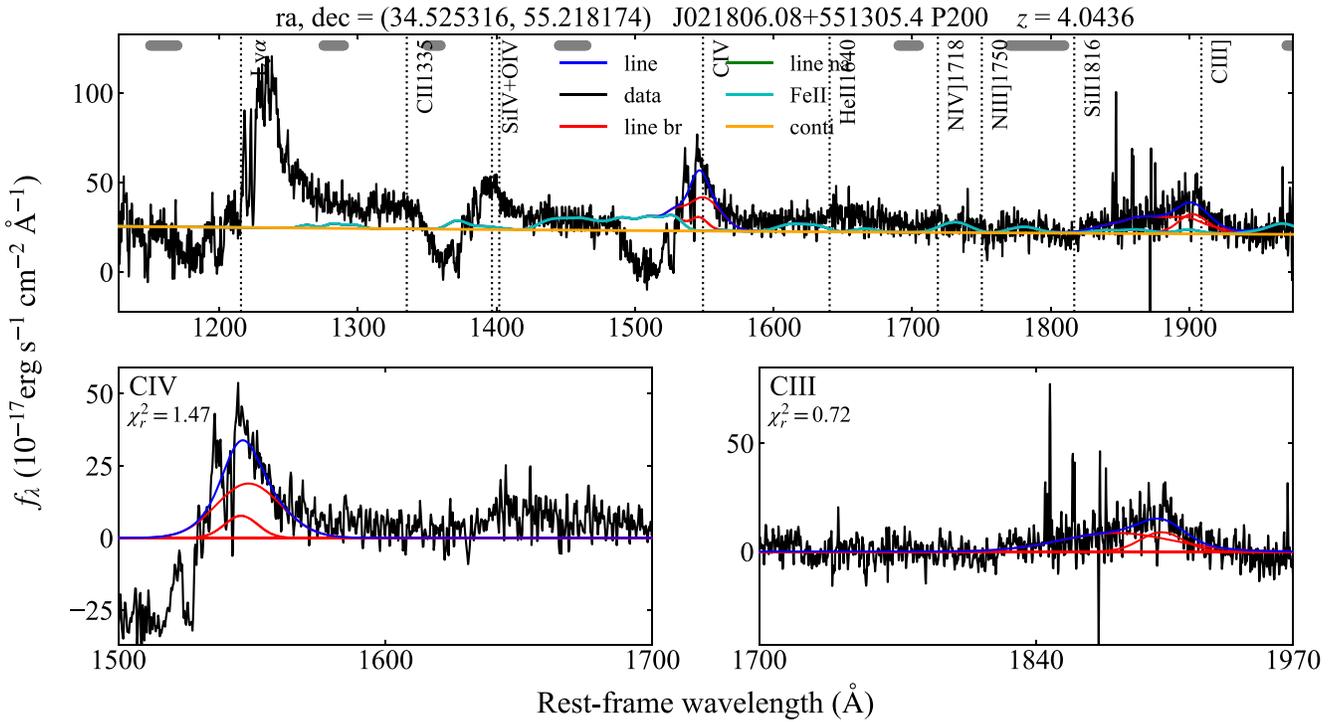

(b)

**Figure 5.** Examples of the spectral fitting for a low-redshift quasar observed by LJT (a) and a high-redshift quasar observed by P200 (b). Black lines denote the total dereddened spectra, yellow lines denote the continua ($f_{pl} + f_{poly}$), cyan lines denote the Fe II templates ($f_{Fe\,II}$), blue lines denote the whole emission lines, red lines denote the broad-line components, and green lines denote the narrow-line components. For a spectrum that is successfully decomposed into a host-galaxy component and a pure quasar component, the purple line denotes the host component, the gray line denotes the pure quasar component, and the pink line denotes the spectrum reconstructed from the PCA host and quasar components.

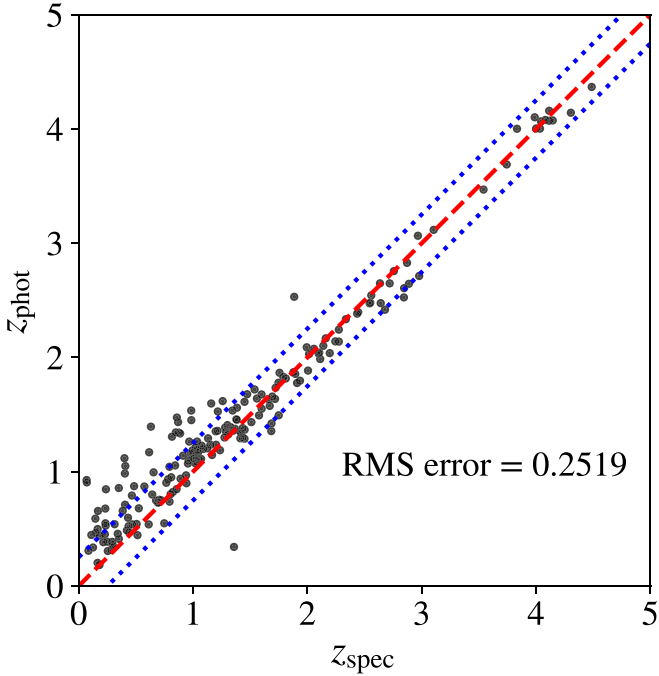

**Figure 6.** Photometric redshift from the GPQ candidate catalog in Paper I against the spectral redshift for the identified GPQs. The red dashed line denotes $z_{\rm phot} = z_{\rm spec}$, and the blue dotted lines mark the margin within one rms error from the red dashed line.

The GPQ spectra are corrected for Galactic extinction with the Planck14 dust map and the extinction law from Wang & Chen (2019). The dereddened spectra are then shifted to the rest frame with the redshifts measured in Section 3.1. After the redshift correction, the spectra are fitted with the multiple components that represent the contributions from the host-galaxy, continuum, and emission lines.

For the spectra of low-redshift quasars, the contribution from the light of their host galaxies can be significant (e.g., Vanden Berk et al. 2001). In order to extract the intrinsic properties of quasars, we decompose each spectrum of quasars with $z < 1.16$ into a host-galaxy component and a quasar component using the principle component analysis (PCA) method (Yip et al. 2004a, 2004b) implemented in PyQSOFit. The redshift limit of $z < 1.16$ corresponds to the two lowest-redshift bins where the host-galaxy eigenspectra are available (Yip et al. 2004b). The PCA host decomposition method assumes that the observed quasar spectrum is a combination of a pure set of eigenspectra (PCA components) of the host galaxy and a pure set of eigenspectra of the quasar. We use 5 eigenspectra of the host galaxy and 20 eigenspectra of the quasar to decompose the spectra. For those spectra that have been successfully decomposed, we remove the host-galaxy components and obtain the pure quasar spectra for the later fitting process. For those spectra that are not decomposed, no subtractions of the host-galaxy components are performed.

We use a power-law model ($f_{\rm pl}$), a three-order polynomial model ($f_{\rm poly}$), and a Fe II model ($f_{\rm Fe\,II}$) to fit a pseudocontinuum ($f_{\rm cont}$) of the quasar spectrum after masking the emission lines:

$$f_{\rm pl} = \beta(\lambda/\lambda_0)^\alpha, \quad (1)$$

$$f_{\rm poly} = \sum_{i=1}^{3} b_i(\lambda - \lambda_1), \quad (2)$$

$$f_{\rm Fe\,II} = c_0 F_{\rm Fe\,II}(\lambda, c_1, c_2), \quad (3)$$

$$f_{\rm cont} = f_{\rm pl} + f_{\rm poly} + f_{\rm Fe\,II}, \quad (4)$$

where $\lambda_0 = 3000$ Å is the reference wavelength, $\beta$ is the amplitude (normalization factor) of the power-law model, $\alpha$ is the power-law index, $b_i$ is the coefficient of the three-order polynomial, and $c_0$, $c_1$, and $c_2$ are the amplitude of the Fe II templates ($F_{\rm Fe\,II}$), FWHM of the Gaussian kernel used to convolve the Fe II templates, and the wavelength shift applied to the Fe II templates, respectively. The optical and UV Fe II templates are built based on Boroson & Green (1992), Vestergaard & Wilkes (2001), Tsuzuki et al. (2006), and Salviander et al. (2007). To prevent the broad absorption lines from being wrongly fitted as continua by the polynomial model, the polynomial component $f_{\rm poly}$ is canceled for quasars at $z \geqslant 2.5$. In addition, by setting "`rej_abs = True`" in the fitting program, we perform one iteration on the continuum fitting to remove the $3\sigma$ outliers below the previous continuum fit for wavelengths <3500 Å, where $\sigma$ is the flux uncertainty of the spectrum. The $3\sigma$ rejection criterion is useful to reduce the impact of the absorption lines on the spectral fitting (Shen et al. 2011, 2019).

The emission-line components are fitted with Gaussian profiles after the continuum component is subtracted from the spectrum. We use the narrow-line components and broad-line components to fit H$\alpha$, H$\beta$, and Mg II, and use only the broad-line components to fit C III] and C IV, as has been suggested by Shen et al. (2011, 2019). The parameters for line fitting are listed in Table 2. To better illustrate the spectral-fitting process described above, a flowchart is shown in Figure 4. Examples of the spectral fitting are shown in Figure 5.

The Monte Carlo (MC) method is applied to estimate the uncertainties of the measured quantities given by QSOFIT-MORE. For each MC trial, the fittings for the continuum and emission lines are performed after a random noise drawn from a Gaussian distribution $\mathcal{N}(0, \sigma^2)$ is added to the quasar spectrum ($\sigma$ being the flux uncertainty). For each spectrum, we calculate the uncertainties of the measured quantities as the standard deviations of the fitting results given by 50 MC trials.

### 3.3. Estimating Virial Black Hole Masses

With the continuum luminosity being a proxy of the broad-line region size (i.e., the $R - L$ relation; e.g., Kaspi et al. 2000; Wu et al. 2004; Bentz et al. 2006; Du et al. 2016), and the broad-line width being a proxy of the virial velocity, we can estimate the single-epoch virial black hole masses ($M_{\rm BH}$) of our GPQ sample using empirical mass-scaling relations that are calibrated with reverberation mapping (RM) masses (e.g., McLure & Dunlop 2004; Vestergaard & Peterson 2006; Wang et al. 2009). To derive the virial black hole masses, we adopt the H$\beta$-based estimator from Vestergaard & Peterson (2006), the Mg II-based estimator from Wang et al. (2009), and the C IV-based estimator from Vestergaard & Peterson (2006) as

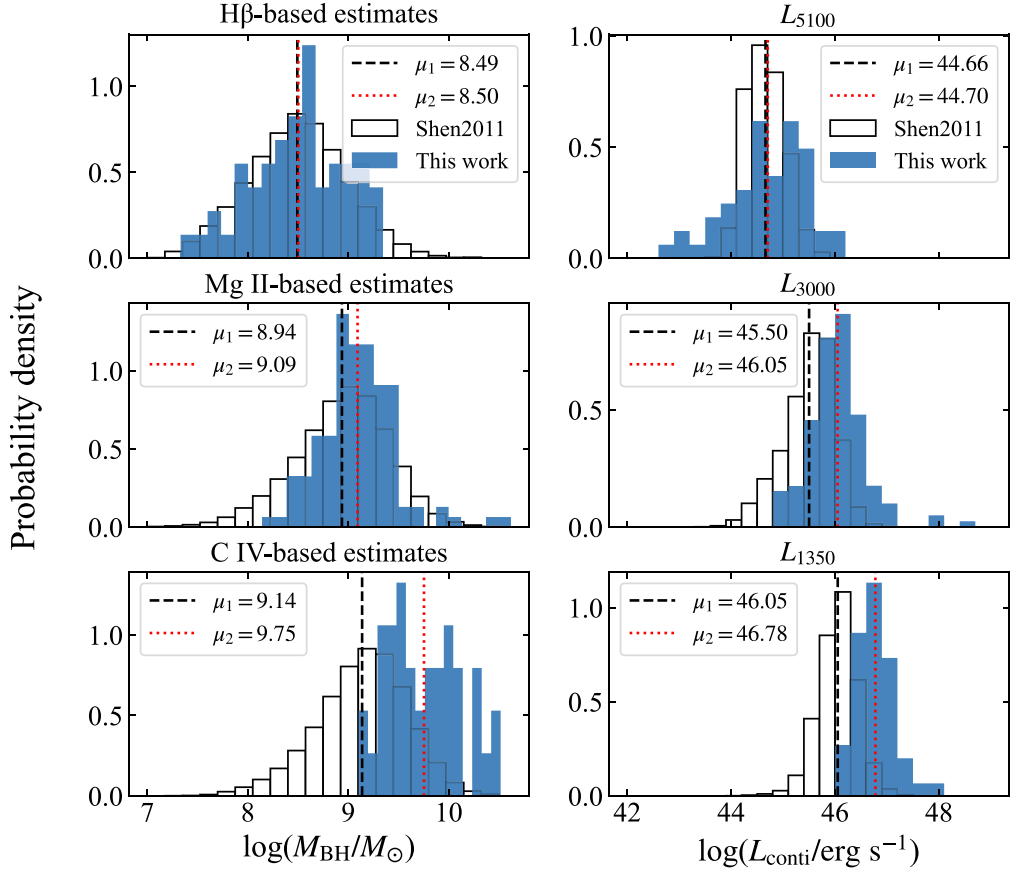

**Figure 7.** Histograms of virial black hole masses (left column) and continuum luminosities (right column) of the SDSS DR7Q sample in Shen et al. (2011) (unfilled bars) and our GPQ sample (filled in blue). The mean values of the DR7Q sample ($\mu_1$) are marked in black dashed lines, while the mean values of our GPQ sample ($\mu_2$) are marked in red dotted lines.

described below:

$$\log(M_{\rm BH}/M_\odot) = \log\left[\left(\frac{{\rm FWHM(H}\beta)}{{\rm km\ s^{-1}}}\right)^2 \left(\frac{L_{5100}}{10^{44}\ {\rm erg\ s^{-1}}}\right)^{0.5}\right] + 0.91, \quad (5)$$

$$\log(M_{\rm BH}/M_\odot) = \log\left[\left(\frac{{\rm FWHM(Mg\ II)}}{{\rm km\ s^{-1}}}\right)^{1.51} \left(\frac{L_{3000}}{10^{44}\ {\rm erg\ s^{-1}}}\right)^{0.5}\right] + 2.60, \quad (6)$$

$$\log(M_{\rm BH}/M_\odot) = \log\left[\left(\frac{{\rm FWHM(C\ IV)}}{{\rm km\ s^{-1}}}\right)^2 \left(\frac{L_{1350}}{10^{44}\ {\rm erg\ s^{-1}}}\right)^{0.53}\right] + 0.66, \quad (7)$$

where the line widths (FWHM) are measured from the broad H$\beta$, broad Mg II, and broad (identical to the whole) C IV lines, and the monochromatic continuum luminosities at 1350, 3000, and 5100 Å ($L_{1350}$, $L_{3000}$, $L_{5100}$) are calculated from the best-fit $f_{\rm pl} + f_{\rm poly}$ components. The measurement uncertainties of the black hole masses are calculated from the propagation of MC errors of the line widths and the monochromatic continuum luminosities. Nevertheless, the absolute uncertainties of the single-epoch black hole mass estimates can be as large as a factor of ~4, because such single-epoch mass estimates deviate from the RM masses by 0.3–0.5 dex (e.g., Vestergaard & Peterson 2006; Wang et al. 2009), and the RM masses themselves are uncertain by a factor of ~3 as compared to the $M_{\rm BH}$–$\sigma_*$ relationship (Onken et al. 2004).

## 4. Results

### 4.1. Validation on the GPQ Candidate Catalog

By matching the GPQ candidates to the SIMBAD (Wenger et al. 2000) database, Paper I has estimated that the purity of quasars on the matched subset is ~90%. Among the 230 observed targets that are found in the GPQ candidate catalog (see Section 2), 196 sources are identified as quasars, 14 sources are identified stars, and 20 sources are labeled as unknown. Therefore the lower limit of the precision of the 230 GPQ candidates is 196/230 = 85.2%, which is close to the aforementioned purity of ~90% on the SIMBAD matches. The lower limit of the fraction of stellar contaminants of the 230 GPQ candidates is 14/230 = 6.1%.

The list of the 17 stellar contaminants and two galaxy contaminants in the 243 observed targets is shown in Table C1. The spectra of the contaminants are displayed in Figure C1. More than half of the stellar contaminants show H$\alpha$ emission lines, which might be raised from emission-line stars or diffuse H$\alpha$-emitting gas (e.g., H II regions, supernova remnants; see Drew et al. 2005; Wu et al. 2021). These stellar and galaxy contaminants have IR excesses and their mid-IR (AllWISE) colors are similar to those of quasars. Those contaminants

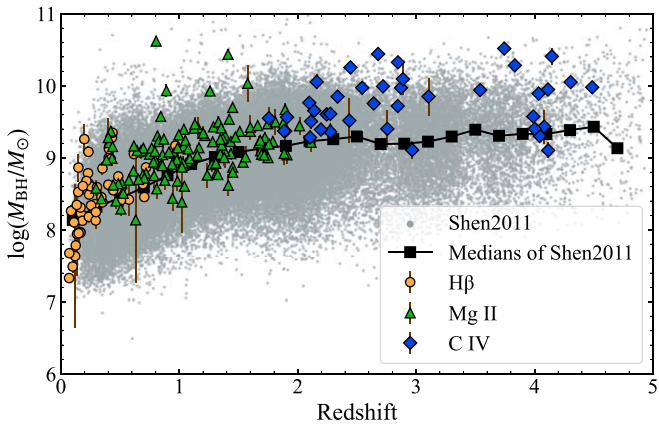

**Figure 8.** Distributions of the black hole masses with redshift of our GPQ sample. Yellow circles represent the H$\beta$-based mass estimates, green triangles represent the Mg II-based mass estimates, and blue diamonds represent the C IV-based mass estimates. Measurement errors derived from the MC errors of spectral fitting are shown as brown error bars. The black hole masses of SDSS DR7Q reported in Shen et al. (2011) are shown as gray dots, and the median black hole masses of DR7Q in 24 redshift bins are shown as black squares.

found in the GPQ candidate catalog also have small (or unmeasured) Gaia proper motions.

The photometric redshifts of the identified GPQs predicted by Paper I are compared with the spectroscopic redshifts in Figure 6. In general, the photometric redshift performance is good, with an rms error of $\sqrt{\sum_{i=1}^{n}(z_{\rm phot}-z_{\rm spec})^2/n} = 0.25$.

Both the high purity of quasars and the small rms error of the photometric redshifts indicate that our GPQ selection methods and the candidate catalog are reliable. Observations on a larger sample of candidates are needed to get a better evaluation result with fewer selection biases toward bright sources (see Section 3.1).

### 4.2. Physical Properties of the 204 Identified GPQs

From the spectral analysis, we obtain 54 estimates for the H$\beta$-based black hole masses ($M_{\rm BH}$), 124 estimates for the Mg II-based $M_{\rm BH}$, and 40 estimates for the C IV-based $M_{\rm BH}$. The spectral-fitting results, black hole mass estimates, and other basic observational properties of the 204 identified GPQs are compiled into a main spectral catalog (Table A1) and an extended spectral catalog (Table A2).

The distributions of the virial black hole masses and the continuum luminosities of our GPQ sample and those of the SDSS DR7Q sample (Shen et al. 2011) are shown in Figure 7. Overall, our GPQ sample has higher virial black hole masses and continuum luminosities than the DR7Q sample does. The mean value of C IV-based $\log(M_{\rm BH})$ of our GPQ sample is larger than that of DR7Q by 0.61 dex, showing the largest deviation of the two samples for all three black hole mass estimates. Similarly, the mean value of $\log(L_{1350})$ of our GPQ sample is larger than that of DR7Q by 0.73 dex, showing the largest deviation of the two samples for all three continuum luminosities.

The distributions of the black hole masses with redshift of our GPQ sample and DR7Q are illustrated in Figure 8. At the low-redshift ($z \lesssim 1.8$) regime, the black hole masses of the GPQ and DR7Q samples are similar, which are based on H$\beta$ and Mg II measurements. At higher redshifts ($z \gtrsim 1.8$), however, most black hole masses of the GPQs are significantly higher than the median values of the DR7Q sample. The excesses of black hole masses and luminosities of GPQs in comparison to DR7Q are due to the selection effect raised from Galactic extinction and the magnitude limit of our current facilities. In the Galactic plane, the intrinsically brighter GPQs are more likely to be detected than the fainter ones. Meanwhile, objects fainter than 20 mag in $i_{P1}$ cannot be covered by the 1 m/2 m telescopes under average weather conditions. Thus we fully expect an artificial difference in the black hole mass distribution between the 204 GPQs and the DR7Q sample.

### 5. Summary and Conclusions

With five optical telescopes, we identified 204 GPQs at $|b| < 20°$ with an average success rate of ~84%, which further proves the high reliability of the GPQ candidate catalog in Paper I. Among the 204 GPQs, 191 are new discoveries. The redshift range of our GPQ sample is $0.069 \leqslant z \leqslant 4.487$, with a distribution peak at $z \approx 1$. From the distributions of both apparent and absolute magnitudes of our GPQs, we can see the identifications so far are biased toward the bright end of the whole GPQ sample.

We perform a multicomponent spectral fitting for all 204 identified GPQs with QSOFITMORE, and obtain the single-epoch virial black hole masses with estimators based on H$\beta$, Mg II, and C IV emission lines. At low redshifts ($z \lesssim 1.8$), our GPQ sample has similar properties to those of SDSS DR7Q. At higher redshifts ($z \gtrsim 1.8$), our GPQ sample has higher black hole masses and continuum luminosities than the DR7Q sample does. Such excesses of black hole masses and luminosities of GPQs are due to the selection effect that intrinsically brighter quasars are more likely to be detected than fainter quasars. The small aperture sizes of our current facilities (except P200) also restrict the identified sample to the bright end.

Our GPQ selection methods and candidate catalog are validated on a subset of 230 observed GPQ candidates, where the lower limit of the precision (purity of quasars) is 85.2%, and the lower limit of the fraction of stellar contaminants is 6.1%. The photometric redshift results are also close to the spectral redshifts of the identified quasars, with an rms error of 0.25.

Most contaminants for our GPQ identifications are stars, more than half of which show H$\alpha$ emission lines, which might be raised from emission-line stars or diffuse H$\alpha$-emitting gas. A minority of the contaminants are galaxies. All the contaminants have IR excesses and their mid-IR colors are similar to those of quasars. Those contaminants found in the GPQ candidate catalog have small (or unmeasured) proper motions as well.

The spectral-fitting results, black hole mass estimates, and other basic observational properties of the 204 identified GPQs are compiled into a main spectral catalog (Table A1), and an extended spectral catalog (Table A2). The data of this paper, including the two spectral catalogs and the spectra of the identified objects (GPQs/contaminants), are available on the China-VO PaperData Repository at doi:10.12149/101140 (version 1).

For future GPQ candidate selections, we will extend our selection methods to more reddened Galactic plane regions with near-IR and mid-IR data, and up-to-date astrometry from Gaia. We will also utilize data from X-ray missions (e.g., eROSITA; Merloni et al. 2012) and radio surveys (e.g., The Karl G. Jansky Very Large Array Sky Survey; Lacy et al. 2020) to determine the most probable GPQ candidates. To identify more GPQs with future observations, we will employ powerful optical telescopes such as LAMOST, which has a high efficiency of spectral acquisition due to its wide field of view (5°) and 4000 optical fibers. In the upcoming Phase II of the LAMOST spectral survey (Zhao et al. 2012; Luo et al. 2015), we expect to discover more GPQs at $|b| \leqslant 20°$. A growing number of GPQs will benefit future astrometric missions and astrophysical programs including Milky Way gas studies.

We thank the support from the National Science Foundation of China (12133001, 11927804, 11721303, 11903003, and 11890693), and the science research grant from the China Manned Space Project with No. CMS-CSST-2021-A06. We thank the referee very much for constructive and helpful suggestions to improve this paper. Y.F. thanks Dr. Anthony G. A. Brown from Leiden Observatory, and Prof. Dr. Sergei A. Klioner from Lohrmann Observatory for their helpful suggestions and support on this project.

We acknowledge the support of the staff of the Xinglong 2.16 m telescope. This work was partially supported by the Open Project Program of the Key Laboratory of Optical Astronomy, National Astronomical Observatories, Chinese Academy of Sciences. We acknowledge the support of the staff of the Lijiang 2.4 m telescope. Funding for the telescope has been provided by Chinese Academy of Sciences and the People's Government of Yunnan Province.

This research uses data obtained through the Telescope Access Program (TAP). Observations obtained with the Hale Telescope at Palomar Observatory were obtained as part of an agreement between the National Astronomical Observatories, Chinese Academy of Sciences, and the California Institute of Technology. This work uses data obtained with the ANU 2.3 m Telescope, Siding Spring Observatory. This work uses data obtained at the MDM Observatory, operated by Dartmouth College, Columbia University, Ohio State University, Ohio University, and the University of Michigan. This research has made use of the SIMBAD database, operated at CDS, Strasbourg, France.

*Facilities:* Beijing:2.16 m, YAO:2.4 m, Hale, McGraw-Hill, ATT .

*Software:* astropy (Astropy Collaboration et al. 2013), ASERA (Yuan et al. 2013), dustmaps (Green 2018), IRAF (Tody 1986, 1993), L.A.Cosmic (van Dokkum 2001; van Dokkum et al. 2012), PyFOSC (Fu 2020), PyQSOFit (Guo et al. 2018), PyRAF (Science Software Branch at STScI 2012), QSOFITMORE (Fu 2021), TOPCAT (Taylor 2005).

## Appendix A
## The Spectral Catalogs of the 204 Identified GPQs

The spectral-fitting results, black hole mass estimates, and other basic observational properties of the 204 GPQs are compiled into a main spectral catalog (Table A1), and an extended spectral catalog with more details of the fitting results. The extended spectral catalog of GPQs contains designation, equatorial and Galactic coordinates, PS1 *i*-band magnitude, line-of-sight $E(B-V)$, redshift, modified Julian date (MJD) of spectroscopic observation, telescope used for the observation, and quantities measured from spectral fitting. The format of the extended spectral catalog is described in Table A2.

Table A1
Main Spectral Catalog of the 204 Identified GPQs

| Designation | $l$ (deg) | $b$ (deg) | $i_{P1}$ (mag) | $z_{VI}$ | log $L_{1350}$ (erg s$^{-1}$) | log $L_{3000}$ (erg s$^{-1}$) | log $L_{5100}$ (erg s$^{-1}$) | log $M_{BH}$(H$\beta$) ($M_\odot$) | log $M_{BH}$(Mg II) ($M_\odot$) | log $M_{BH}$(C IV) ($M_\odot$) | Telescope |
|---|---|---|---|---|---|---|---|---|---|---|---|
| J000127.44+462139.0 | 114.0119664 | −15.6434617 | 16.941 | 1.4343 | | 46.371 ± 0.004 | | | 9.058 ± 0.112 | | XLT |
| J000627.25+542823.2 | 116.3635814 | −7.8239422 | 16.958 | 2.3368 | 47.48 ± 0.012 | | | | | 9.849 ± 0.056 | MDM13 |
| J001000.04+490639.3 | 115.9875894 | −13.2013190 | 16.768 | 0.4078 | | 45.862 ± 0.002 | 45.067 ± 0.007 | 9.159 ± 0.026 | 9.21 ± 0.029 | | LJT |
| J001554.18+560257.5 | 117.9523862 | −6.4784876 | 16.809 | 0.1684 | | | 44.633 ± 0.006 | 8.229 ± 0.067 | | | LJT |
| J002801.07+541121.6 | 119.4717617 | −8.5249846 | 18.421 | 2.7569 | 46.786 ± 0.037 | | | | | 9.401 ± 0.339 | LJT |
| J004923.42+451358.0 | 122.5536527 | −17.6374529 | 19.212 | 4.1152 | 46.497 ± 0.011 | | | | | 9.103 ± 0.094 | P200 |
| J005158.90+542241.5 | 123.0119858 | −8.4934794 | 17.480 | 1.2591 | | 46.5 ± 0.001 | | | 9.927 ± 0.001 | | LJT |
| J011257.00+560721.2 | 125.9468306 | −6.6232184 | 18.673 | 2.8720 | 46.968 ± 0.008 | | | | | 9.972 ± 0.105 | P200 |
| J011927.16+460539.8 | 127.9914015 | −16.5019679 | 16.571 | 0.7945 | | 46.299 ± 0.002 | | | 8.972 ± 0.021 | | LJT |
| J012158.13+531032.2 | 127.5605980 | −9.4215318 | 16.139 | 1.0382 | | 46.573 ± 0.002 | | | 9.396 ± 0.022 | | LJT |
| J013716.53+542146.4 | 129.6433750 | −7.9104611 | 18.394 | 1.7222 | 46.195 ± ?[a] | 45.989 ± 0.009 | | | 9.28 ± 0.093 | | LJT |

**Note.** This table is published in its entirety in the machine-readable format, with unmeasurable values indicated with −999. Only a portion of the table is displayed here. A copy in FITS format can be found in the China-VO PaperData Repository at doi:10.12149/101140 (version 1) .

[a] A question mark ("?") as a substitute for the measurement error of log$L_{1350}$ indicates that the measurement error cannot be determined because the rest frame 1350 Å is close to the edge of the spectrum. In such cases, black hole masses based on C IV are not reported.

(This table is available in its entirety in machine-readable form.)

Table A2
Format of the Extended Spectral Catalog of the 204 Identified GPQs

| Column | Name | Format | Unit | Description |
|---|---|---|---|---|
| 1 | DESIGNATION | STRING | | Object designation hhmmss.ss+ddmmss.s (J2000) based on PS1 coordinates |
| 2 | R.A. | DOUBLE | deg | PS1 R.A. in decimal degrees (J2000) (weighted mean) at mean epoch |
| 3 | Decl. | DOUBLE | deg | PS1 decl. in decimal degrees (J2000) (weighted mean) at mean epoch |
| 4 | GLON | DOUBLE | deg | Galactic longitude in decimal degrees |
| 5 | GLAT | DOUBLE | deg | Galactic latitude in decimal degrees |
| 6 | IMAG | DOUBLE | mag | Mean point-spread function AB magnitude from PS1 $i$-filter detections |
| 7 | EBV | DOUBLE | mag | Line-of-sight $E(B-V)$ given by the Planck14 dust map |
| 8 | Z_VI | DOUBLE | | Redshift from visual inspection |
| 9 | MI_Z2 | DOUBLE | mag | SDSS $i$-band absolute magnitude $M_i(z=2)$, $K$-corrected to $z=2$ following Richards et al. (2006) |
| 10 | MJD | DOUBLE | day | Modified Julian date of spectroscopic observation |
| 11 | TELESCOPE | STRING | | Telescope used to acquire the spectrum |
| 12 | SNR_SPEC | DOUBLE | | Median signal-to-noise ratio (S/N) per pixel of the spectrum |
| 13 | LOGL1350 | DOUBLE | erg s$^{-1}$ | Logarithmic continuum luminosity at rest frame 1350 Å |
| 14 | ERR_LOGL1350 | DOUBLE | erg s$^{-1}$ | Measurement error in LOGL1350 |
| 15 | LOGL3000 | DOUBLE | erg s$^{-1}$ | Logarithmic continuum luminosity at rest frame 3000 Å |
| 16 | ERR_LOGL3000 | DOUBLE | erg s$^{-1}$ | Measurement error in LOGL3000 |
| 17 | LOGL5100 | DOUBLE | erg s$^{-1}$ | Logarithmic continuum luminosity at rest frame 5100 Å |
| 18 | ERR_LOGL5100 | DOUBLE | erg s$^{-1}$ | Measurement error in LOGL5100 |
| 19 | LOGBH_HB | DOUBLE | $M_\odot$ | Logarithmic single-epoch virial black hole (BH) mass based on H$\beta$ |
| 20 | ERR_LOGBH_HB | DOUBLE | $M_\odot$ | Measurement error in LOGBH_HB |
| 21 | LOGBH_MGII | DOUBLE | $M_\odot$ | Logarithmic single-epoch virial BH mass based on Mg II |
| 22 | ERR_LOGBH_MGII | DOUBLE | $M_\odot$ | Measurement error in LOGBH_MgII |
| 23 | LOGBH_CIV | DOUBLE | $M_\odot$ | Logarithmic single-epoch virial BH mass based on C IV |
| 24 | ERR_LOGBH_CIV | DOUBLE | $M_\odot$ | Measurement error in LOGBH_CIV |
| 25 | FE_UV_NORM | DOUBLE | | The normalization factor of the ultraviolet Fe II template |
| 26 | FE_UV_SHIFT | DOUBLE | | The wavelength shift of the the ultraviolet Fe II template |
| 27 | FE_UV_FWHM | DOUBLE | | The Gaussian FWHM applied to convolve the ultraviolet Fe II template |
| 28 | FE_OP_NORM | DOUBLE | | The normalization factor of the optical Fe II template |
| 29 | FE_OP_SHIFT | DOUBLE | | The wavelength shift of the optical Fe II template |
| 30 | FE_OP_FWHM | DOUBLE | | The Gaussian FWHM applied to convolve the optical Fe II template |
| 31 | PL_NORM | DOUBLE | $10^{-17} \times$ erg s$^{-1}$ cm$^{-2}$ Å$^{-1}$ | The normalization factor of the power-law model |
| 32 | PL_ALPHA | DOUBLE | | Wavelength power-law index |
| 33 | POLY_A | DOUBLE | | Coefficient of the three-order polynomial model ($b_1$) |
| 34 | POLY_B | DOUBLE | | Coefficient of the three-order polynomial model ($b_2$) |
| 35 | POLY_C | DOUBLE | | Coefficient of the three-order polynomial model ($b_3$) |
| 36 | LINE_NPIX_CIV | LONG | | Number of good pixels for the rest frame 1500–1700 Å |
| 37 | LINE_STATUS_CIV | LONG | | Line-fitting status of C IV |
| 38 | LINE_MED_SN_CIV | DOUBLE | | Median S/N per pixel for the rest frame 1500–1700 Å |
| 39 | FWHM_CIV | DOUBLE | km s$^{-1}$ | FWHM of the whole C IV |
| 40 | ERR_FWHM_CIV | DOUBLE | km s$^{-1}$ | Measurement error in FWHM_CIV |
| 41 | EW_CIV | DOUBLE | Å | Rest-frame equivalent width of the whole C IV |
| 42 | ERR_EW_CIV | DOUBLE | Å | Measurement error in EW_CIV |
| 43 | FLUX_CIV | DOUBLE | $10^{-17}$ erg s$^{-1}$ cm$^{-2}$ | Flux of the whole C IV |
| 44 | ERR_FLUX_CIV | DOUBLE | $10^{-17}$ erg s$^{-1}$ cm$^{-2}$ | Measurement error in FLUX_CIV |
| 45 | LINE_NPIX_CIII | LONG | | Number of good pixels for the rest frame 1700–1970 Å |
| 46 | LINE_STATUS_CIII | LONG | | Line-fitting status of CIII |
| 47 | LINE_MED_SN_CIII | DOUBLE | | Median S/N per pixel for the rest frame 1700–1970 Å |
| 48 | FWHM_CIII | DOUBLE | km s$^{-1}$ | FWHM of the whole CIII |
| 49 | ERR_FWHM_CIII | DOUBLE | km s$^{-1}$ | Measurement error in FWHM_CIII |
| 50 | EW_CIII | DOUBLE | Å | Rest-frame equivalent width of the whole CIII |
| 51 | ERR_EW_CIII | DOUBLE | Å | Measurement error in EW_CIII |
| 52 | FLUX_CIII | DOUBLE | $10^{-17}$ erg s$^{-1}$ cm$^{-2}$ | Flux of the whole CIII |
| 53 | ERR_FLUX_CIII | DOUBLE | $10^{-17}$ erg s$^{-1}$ cm$^{-2}$ | Measurement error in FLUX_CIII |
| 54 | LINE_NPIX_MGII | LONG | | Number of good pixels for the rest frame 2700–2900 Å |
| 55 | LINE_STATUS_MGII | LONG | | Line-fitting status of Mg II |
| 56 | LINE_MED_SN_MGII | DOUBLE | | Median S/N per pixel for the rest frame 2700–2900 Å |
| 57 | FWHM_BROAD_MGII | DOUBLE | km s$^{-1}$ | FWHM of the whole broad Mg II |
| 58 | ERR_FWHM_BROAD_MGII | DOUBLE | km s$^{-1}$ | Measurement error in FWHM_BROAD_MGII |
| 59 | EW_BROAD_MGII | DOUBLE | Å | Rest-frame equivalent width of the whole broad Mg II |
| 60 | ERR_EW_BROAD_MGII | DOUBLE | Å | Measurement error in EW_BROAD_MGII |
| 61 | FLUX_BROAD_MGII | DOUBLE | $10^{-17}$ erg s$^{-1}$ cm$^{-2}$ | Flux of the whole broad Mg II |

Table A2
(Continued)

| Column | Name | Format | Unit | Description |
|---|---|---|---|---|
| 62 | ERR_FLUX_BROAD_MGII | DOUBLE | $10^{-17}$ erg s$^{-1}$ cm$^{-2}$ | Measurement error in FLUX_BROAD_MGII |
| 63 | FWHM_NARROW_MGII | DOUBLE | km s$^{-1}$ | FWHM of the narrow Mg II |
| 64 | ERR_FWHM_NARROW_MGII | DOUBLE | km s$^{-1}$ | Measurement error in FWHM_NARROW_MGII |
| 65 | EW_NARROW_MGII | DOUBLE | Å | Rest-frame equivalent width of the narrow Mg II |
| 66 | ERR_EW_NARROW_MGII | DOUBLE | Å | Measurement error in EW_NARROW_MGII |
| 67 | FLUX_NARROW_MGII | DOUBLE | $10^{-17}$ erg s$^{-1}$ cm$^{-2}$ | Flux of the narrow Mg II |
| 68 | ERR_FLUX_NARROW_MGII | DOUBLE | $10^{-17}$ erg s$^{-1}$ cm$^{-2}$ | Measurement error in FLUX_NARROW_MGII |
| 69 | LINE_NPIX_HB | LONG | | Number of good pixels for the rest frame 4640-5100 Å |
| 70 | LINE_STATUS_HB | LONG | | Line-fitting status of H$\beta$ |
| 71 | LINE_MED_SN_HB | DOUBLE | | Median S/N per pixel for the rest frame 4640-5100 Å |
| 72 | FWHM_BROAD_HB | DOUBLE | km s$^{-1}$ | FWHM of broad H$\beta$ |
| 73 | ERR_FWHM_BROAD_HB | DOUBLE | km s$^{-1}$ | Measurement error in FWHM_BROAD_HB |
| 74 | EW_BROAD_HB | DOUBLE | Å | Rest-frame equivalent width of broad H$\beta$ |
| 75 | ERR_EW_BROAD_HB | DOUBLE | Å | Measurement error in EW_BROAD_HB |
| 76 | FLUX_BROAD_HB | DOUBLE | $10^{-17}$ erg s$^{-1}$ cm$^{-2}$ | Flux of broad H$\beta$ |
| 77 | ERR_FLUX_BROAD_HB | DOUBLE | $10^{-17}$ erg s$^{-1}$ cm$^{-2}$ | Measurement error in FLUX_BROAD_HB |
| 78 | FWHM_NARROW_HB | DOUBLE | km s$^{-1}$ | FWHM of narrow H$\beta$ |
| 79 | ERR_FWHM_NARROW_HB | DOUBLE | km s$^{-1}$ | Measurement error in FWHM_NARROW_HB |
| 80 | EW_NARROW_HB | DOUBLE | Å | Rest-frame equivalent width of narrow H$\beta$ |
| 81 | ERR_EW_NARROW_HB | DOUBLE | Å | Measurement error in EW_NARROW_HB |
| 82 | FLUX_NARROW_HB | DOUBLE | $10^{-17}$ erg s$^{-1}$ cm$^{-2}$ | Flux of narrow H$\beta$ |
| 83 | ERR_FLUX_NARROW_HB | DOUBLE | $10^{-17}$ erg s$^{-1}$ cm$^{-2}$ | Measurement error in FLUX_NARROW_HB |
| 84 | FWHM_OIII_4959 | DOUBLE | km s$^{-1}$ | FWHM of [O III]$\lambda$4959 |
| 85 | ERR_FWHM_OIII_4959 | DOUBLE | km s$^{-1}$ | Measurement error in FWHM_OIII_4959 |
| 86 | EW_OIII_4959 | DOUBLE | Å | Rest-frame equivalent width of [O III]$\lambda$4959 |
| 87 | ERR_EW_OIII_4959 | DOUBLE | Å | Measurement error in EW_OIII_4959 |
| 88 | FLUX_OIII_4959 | DOUBLE | $10^{-17}$ erg s$^{-1}$ cm$^{-2}$ | Flux of [O III]$\lambda$4959 |
| 89 | ERR_FLUX_OIII_4959 | DOUBLE | $10^{-17}$ erg s$^{-1}$ cm$^{-2}$ | Measurement error in FLUX_OIII_4959 |
| 90 | FWHM_OIII_5007 | DOUBLE | km s$^{-1}$ | FWHM of [O III]$\lambda$5007 |
| 91 | ERR_FWHM_OIII_5007 | DOUBLE | km s$^{-1}$ | Measurement error in FWHM_OIII_5007 |
| 92 | EW_OIII_5007 | DOUBLE | Å | Rest-frame equivalent width of [O III]$\lambda$5007 |
| 93 | ERR_EW_OIII_5007 | DOUBLE | Å | Measurement error in EW_OIII_5007 |
| 94 | FLUX_OIII_5007 | DOUBLE | $10^{-17}$ erg s$^{-1}$ cm$^{-2}$ | Flux of [O III]$\lambda$5007 |
| 95 | ERR_FLUX_OIII_5007 | DOUBLE | $10^{-17}$ erg s$^{-1}$ cm$^{-2}$ | Measurement error in FLUX_OIII_5007 |
| 96 | LINE_NPIX_HA | LONG | | Number of good pixels for the rest frame 6400-6800 Å |
| 97 | LINE_STATUS_HA | LONG | | Line-fitting status of H$\alpha$ |
| 98 | LINE_MED_SN_HA | DOUBLE | | Median S/N per pixel for the rest frame 6400-6800 Å |
| 99 | FWHM_BROAD_HA | DOUBLE | km s$^{-1}$ | FWHM of broad H$\alpha$ |
| 100 | ERR_FWHM_BROAD_HA | DOUBLE | km s$^{-1}$ | Measurement error in FWHM_BROAD_HA |
| 101 | EW_BROAD_HA | DOUBLE | Å | Rest-frame equivalent width of broad H$\alpha$ |
| 102 | ERR_EW_BROAD_HA | DOUBLE | Å | Measurement error in EW_BROAD_HA |
| 103 | FLUX_BROAD_HA | DOUBLE | $10^{-17}$ erg s$^{-1}$ cm$^{-2}$ | Flux of broad H$\alpha$ |
| 104 | ERR_FLUX_BROAD_HA | DOUBLE | $10^{-17}$ erg s$^{-1}$ cm$^{-2}$ | Measurement error in FLUX_BROAD_HA |
| 105 | FWHM_NARROW_HA | DOUBLE | km s$^{-1}$ | FWHM of narrow H$\alpha$ |
| 106 | ERR_FWHM_NARROW_HA | DOUBLE | km s$^{-1}$ | Measurement error in FWHM_NARROW_HA |
| 107 | EW_NARROW_HA | DOUBLE | Å | Rest-frame equivalent width of narrow H$\alpha$ |
| 108 | ERR_EW_NARROW_HA | DOUBLE | Å | Measurement error in EW_NARROW_HA |
| 109 | FLUX_NARROW_HA | DOUBLE | $10^{-17}$ erg s$^{-1}$ cm$^{-2}$ | Flux of narrow H$\alpha$ |
| 110 | ERR_FLUX_NARROW_HA | DOUBLE | $10^{-17}$ erg s$^{-1}$ cm$^{-2}$ | Measurement error in FLUX_NARROW_HA |
| 111 | FWHM_NII_6549 | DOUBLE | km s$^{-1}$ | FWHM of [N II]$\lambda$6549 |
| 112 | ERR_FWHM_NII_6549 | DOUBLE | km s$^{-1}$ | Measurement error in FWHM_NII_6549 |
| 113 | EW_NII_6549 | DOUBLE | Å | Rest-frame equivalent width of [N II]$\lambda$6549 |
| 114 | ERR_EW_NII_6549 | DOUBLE | Å | Measurement error in EW_NII_6549 |
| 115 | FLUX_NII_6549 | DOUBLE | $10^{-17}$ erg s$^{-1}$ cm$^{-2}$ | Flux of [N II]$\lambda$6549 |
| 116 | ERR_FLUX_NII_6549 | DOUBLE | $10^{-17}$ erg s$^{-1}$ cm$^{-2}$ | Measurement error in FLUX_NII_6549 |
| 117 | FWHM_NII_6585 | DOUBLE | km s$^{-1}$ | FWHM of [N II]$\lambda$6585 |
| 118 | ERR_FWHM_NII_6585 | DOUBLE | km s$^{-1}$ | Measurement error in FWHM_NII_6585 |
| 119 | EW_NII_6585 | DOUBLE | Å | Rest-frame equivalent width of [N II]$\lambda$6585 |
| 120 | ERR_EW_NII_6585 | DOUBLE | Å | Measurement error in EW_NII_6585 |
| 121 | FLUX_NII_6585 | DOUBLE | $10^{-17}$ erg s$^{-1}$ cm$^{-2}$ | Flux of [N II]$\lambda$6585 |
| 122 | ERR_FLUX_NII_6585 | DOUBLE | $10^{-17}$ erg s$^{-1}$ cm$^{-2}$ | Measurement error in FLUX_NII_6585 |
| 123 | FWHM_SII_6718 | DOUBLE | km s$^{-1}$ | FWHM of [S II]$\lambda$6718 |
| 124 | ERR_FWHM_SII_6718 | DOUBLE | km s$^{-1}$ | Measurement error in FWHM_SII_6718 |
| 125 | EW_SII_6718 | DOUBLE | Å | Rest-frame equivalent width of [S II]$\lambda$6718 |

Table A2
(Continued)

| Column | Name | Format | Unit | Description |
| --- | --- | --- | --- | --- |
| 126 | ERR_EW_SII_6718 | DOUBLE | Å | Measurement error in EW_SII_6718 |
| 127 | FLUX_SII_6718 | DOUBLE | $10^{-17}$ erg s$^{-1}$ cm$^{-2}$ | Flux of [S II]$\lambda$6718 |
| 128 | ERR_FLUX_SII_6718 | DOUBLE | $10^{-17}$ erg s$^{-1}$ cm$^{-2}$ | Measurement error in FLUX_SII_6718 |
| 129 | FWHM_SII_6732 | DOUBLE | km s$^{-1}$ | FWHM of [S II]$\lambda$6732 |
| 130 | ERR_FWHM_SII_6732 | DOUBLE | km s$^{-1}$ | Measurement error in FWHM_SII_6732 |
| 131 | EW_SII_6732 | DOUBLE | Å | Rest-frame equivalent width of [S II]$\lambda$6732 |
| 132 | ERR_EW_SII_6732 | DOUBLE | Å | Measurement error in EW_SII_6732 |
| 133 | FLUX_SII_6732 | DOUBLE | $10^{-17}$ erg s$^{-1}$ cm$^{-2}$ | Flux of [S II]$\lambda$6732 |
| 134 | ERR_FLUX_SII_6732 | DOUBLE | $10^{-17}$ erg s$^{-1}$ cm$^{-2}$ | Measurement error in FLUX_SII_6732 |

**Note.** This table is published in its entirety in a machine-readable format, with unmeasurable values indicated with −999. Only a portion of the table is displayed here. A copy in FITS format can be found in the China-VO PaperData Repository at doi:10.12149/101140 (version 1).

(This table is available in its entirety in machine-readable form.)

# Appendix B
# The Spectra of the 204 Identified GPQs

The spectra (uncorrected for Galactic extinction) of the 204 identified GPQs are plotted in Figure 9. The spectra are recorded in FITS format and are zipped into a single file, which is available on the China-VO PaperData Repository at doi:10.12149/101140 (version 1).

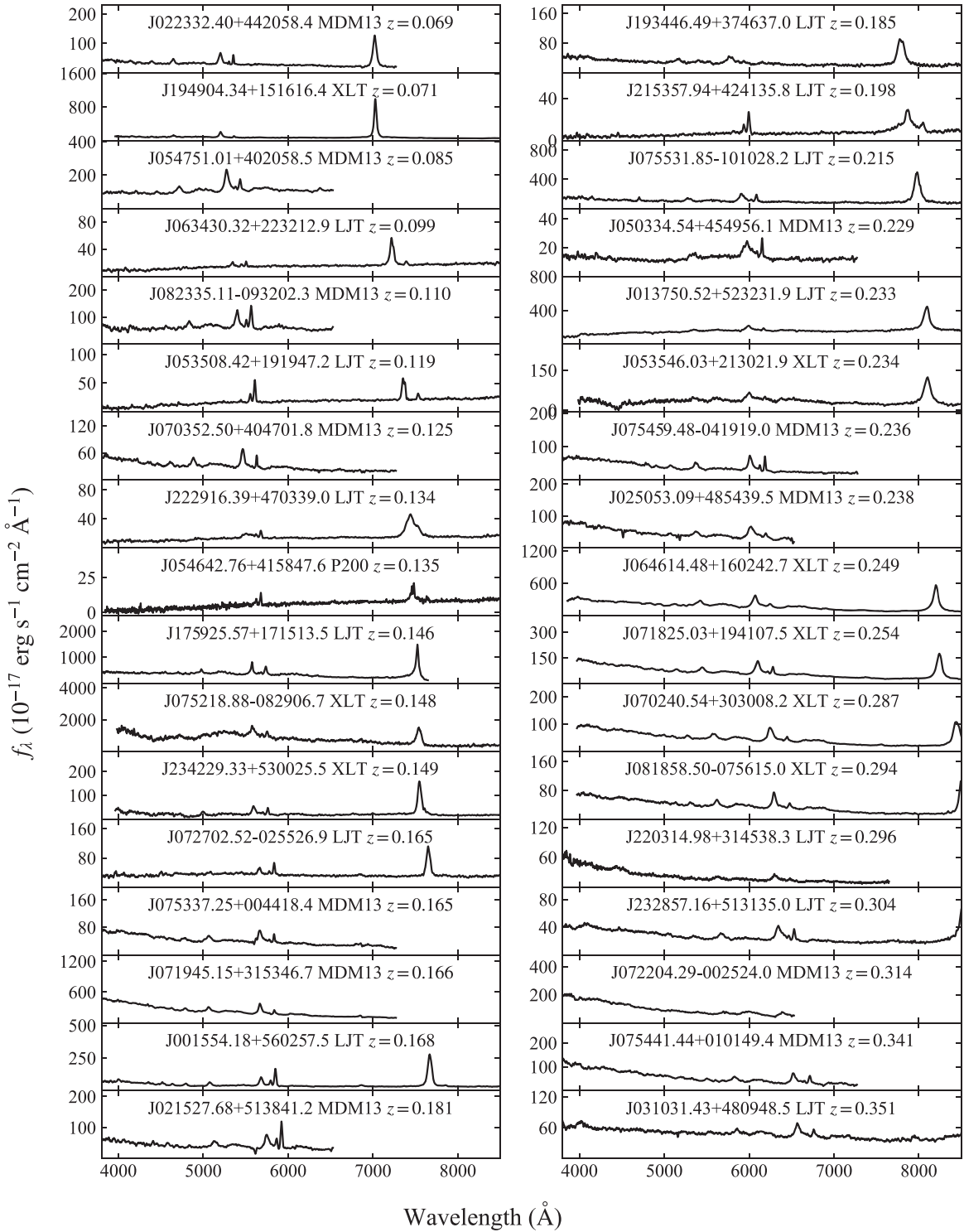

**Figure 9.** The spectra of GPQs identified in this work in the observed frame, which are smoothed for better visualizations. In each column, the quasars are arranged with ascending redshift values. The 204 GPQs are divided into six portions with ascending redshift values. This figure shows the first portion of the spectra of GPQs. These spectra in FITS format are available via the China-VO PaperData Repository at doi:10.12149/101140 (version 1).

(The complete figure set (6 images) is available.)

# Appendix C
# The List and Spectra of the Contaminants

The list of the 19 stellar and galaxy contaminants (17 stars and two galaxies) is shown as Table C1. The spectra (uncorrected for Galactic extinction) of the contaminants are plotted in Figure C1. The spectra are recorded in FITS format and are zipped into a single file, which is available on the China-VO PaperData Repository at doi:10.12149/101140 (version 1).

**Table C1**
List of the Stellar and Galaxy Contaminants

| Designation | R.A. (deg) | Decl. (deg) | $l$ (deg) | $b$ (deg) | $i_{P1}$ (mag) | Class | $z_{VI}$ | MJD | Telescope |
|---|---|---|---|---|---|---|---|---|---|
| J000916.81+601212.6 | 2.3200242 | 60.2035106 | 117.7129394 | −2.2403495 | 15.938 | star | | 58,844.427 | XLT |
| J023942.68+620045.9 | 39.9278364 | 62.0127512 | 135.2680397 | 1.7624536 | 17.362 | star | | 58,813.64 | LJT |
| J024710.69+560037.4 | 41.7945413 | 56.0103828 | 138.6510022 | −3.2882111 | 15.128 | star | | 58,844.539 | XLT |
| J040600.17+562248.2 | 61.500693 | 56.3800521 | 147.5265484 | 3.1013363 | 18.159 | star | | 59,174.196 | P200 |
| J045454.70+372725.0 | 73.7279116 | 37.4569403 | 166.9543974 | −3.819519 | 17.924 | star | | 59,174.3 | P200 |
| J050011.58+475140.8 | 75.0482587 | 47.8613468 | 159.4076651 | 3.4129404 | 15.959 | star | | 58,843.632 | XLT |
| J050729.64+372004.2 | 76.8734943 | 37.3345034 | 168.5854225 | −1.9204993 | 18.201 | star | | 59,174.308 | P200 |
| J051824.97+502341.4 | 79.6040435 | 50.3948271 | 159.1247101 | 7.3137911 | 19.595 | star | | 59,173.344 | P200 |
| J052201.85+321907.5 | 80.5076881 | 32.3187437 | 174.4027242 | −2.4210279 | 17.712 | star | | 59,174.25 | P200 |
| J053021.04+381559.7 | 82.5876593 | 38.2665872 | 170.4172725 | 2.3094733 | 18.265 | star | | 59,174.232 | P200 |
| J053812.11+273607.2 | 84.550475 | 27.6019906 | 180.2717419 | −2.0913483 | 15.16 | star | | 59,229.232 | MDM13 |
| J054914.23+321426.0 | 87.3093056 | 32.2405518 | 177.5743624 | 2.3757573 | 18.537 | star | | 59,174.347 | P200 |
| J055704.62+600026.0 | 89.2692609 | 60.0072102 | 153.4476105 | 16.8808308 | 19.644 | star | | 59,173.463 | P200 |
| J055800.12+472601.0 | 89.5005101 | 47.4336145 | 165.0961942 | 11.3458206 | 16.851 | star | | 58,521.59 | LJT |
| J061919.41+150218.0 | 94.8308627 | 15.0383337 | 195.9286423 | −0.0993244 | 16.495 | star | | 59,171.404 | MDM13 |
| J064439.58+060638.3 | 101.1649098 | 6.1106464 | 206.7219312 | 1.2818258 | 16.77 | star | | 59,227.297 | MDM13 |
| J195628.70+454857.9 | 299.1195684 | 45.8160857 | 80.3152638 | 8.8187372 | 17.585 | star | | 58,252.855 | LJT |
| J052008.51+330019.1 | 80.0354757 | 33.0052991 | 173.6112474 | −2.3574211 | 17.912 | galaxy | 0.0463 | 58,496.535 | XLT |
| J072238.30+042844.4 | 110.6595775 | 4.4789981 | 212.479093 | 8.9763836 | 17.983 | galaxy | 0.1038 | 58,496.738 | XLT |

**Note.** This table is published in its entirety in a machine-readable format, with redshift values of stars filled with 0. A copy in FITS format can be found in the China-VO PaperData Repository at doi:10.12149/101140 (version 1).

(This table is available in machine-readable form.)

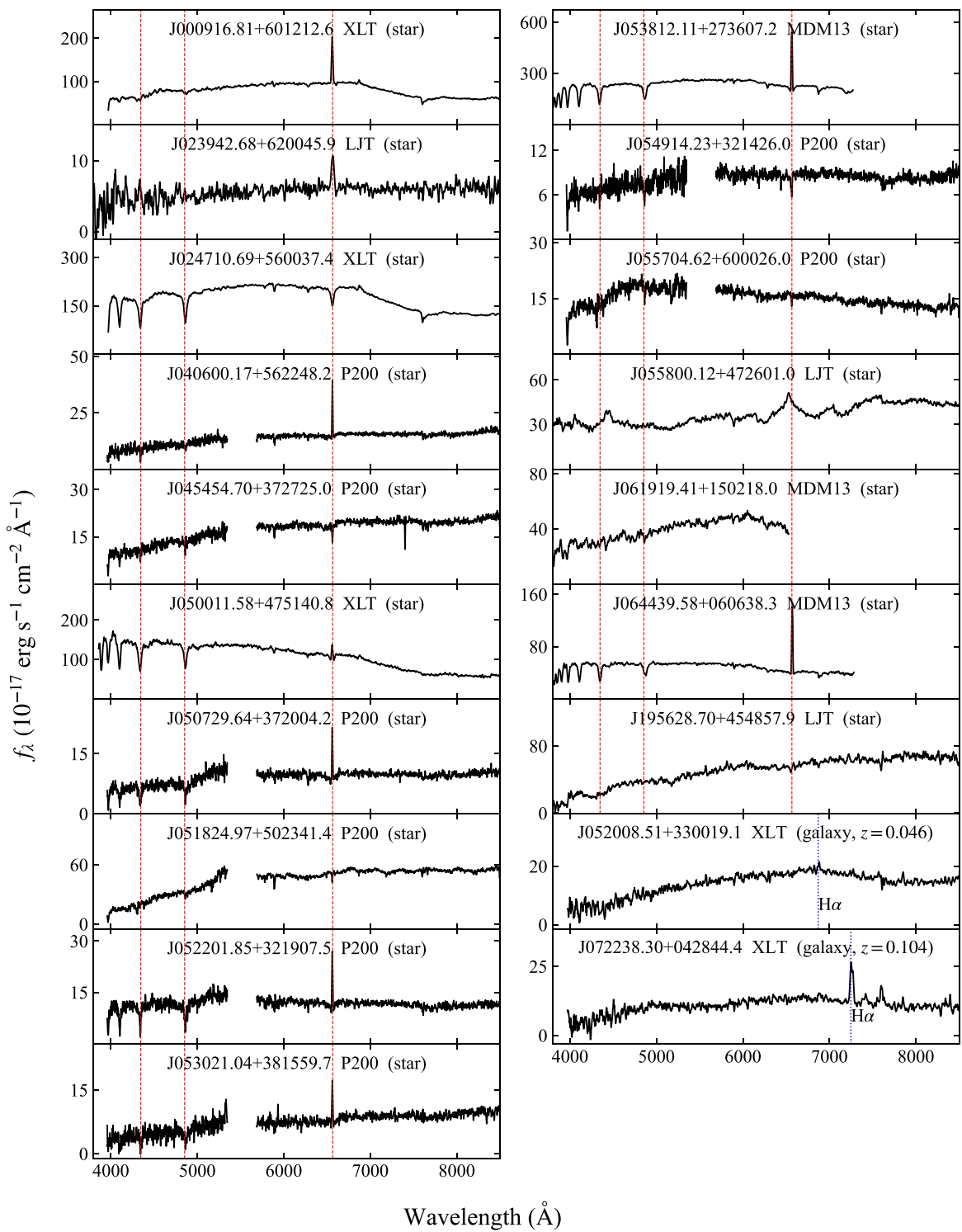

**Figure C1.** The spectra of the stellar and galaxy contaminants. For the spectra of stellar contaminants, red dashed lines mark the wavelengths of the H$\alpha$ (6563 Å), H$\beta$ (4861 Å), and H$\gamma$ (4341 Å) emission/absorption lines. For the spectra of galaxies, blue dotted lines mark the central wavelengths of the redshifted H$\alpha$ emission lines.

**ORCID iDs**

Yuming Fu ● https://orcid.org/0000-0002-0759-0504
Xue-Bing Wu ● https://orcid.org/0000-0002-7350-6913
Linhua Jiang ● https://orcid.org/0000-0003-4176-6486
Yanxia Zhang ● https://orcid.org/0000-0002-6610-5265
Y. L. Ai ● https://orcid.org/0000-0001-9312-4640
Qian Yang ● https://orcid.org/0000-0002-6893-3742
Qinchun Ma ● https://orcid.org/0000-0003-0827-2273
Xiaotong Feng ● https://orcid.org/0000-0003-0174-5920
Ravi Joshi ● https://orcid.org/0000-0002-5535-4186
Christian Wolf ● https://orcid.org/0000-0002-4569-016X
Jiang-Tao Li ● https://orcid.org/0000-0001-6239-3821

Jun-Jie Jin ⓘ https://orcid.org/0000-0002-8402-3722
Su Yao ⓘ https://orcid.org/0000-0002-9728-1552
Jian-Guo Wang ⓘ https://orcid.org/0000-0003-4156-3793
Kai-Xing Lu ⓘ https://orcid.org/0000-0002-2310-0982
Jie Zheng ⓘ https://orcid.org/0000-0001-6637-6973
Pei Zuo ⓘ https://orcid.org/0000-0003-3948-9192